\documentclass[journal]{IEEEtran}
\ifCLASSINFOpdf
\else
\fi
\usepackage[utf8]{inputenc}
\usepackage{cite}
\usepackage{caption}
\usepackage{amsmath}   
\usepackage{svg}
\usepackage{amssymb}   
\usepackage{graphicx}  
\usepackage{caption}   
\usepackage{subfig}    
\usepackage{textgreek} 
\newcommand{\angstrom}{\textup{\AA}} 
\usepackage{stfloats}  
\usepackage{booktabs}      
\usepackage{url}  
\usepackage{threeparttable}
\usepackage{array}         
\usepackage{amsmath}
\usepackage{booktabs}      
\usepackage{caption}       
\usepackage{threeparttable}
\usepackage{multirow}      
\usepackage{array} 

\begin{document}

\title{MP-GCAN: a highly accurate classifier for $\alpha$-helical membrane proteins and $\beta$-barrel proteins}
\author{Kunyang Li, Hongfu Lou, Dinan Peng}

\maketitle

\begin{abstract}
Abstract
Membrane protein classification is a fundamental task in structural bioinformatics, critical to understanding protein functions and accelerating drug discovery. In this study, we propose MP-GCAN, a novel graph-based classification model that leverages both spatial and sequential features of proteins. MP-GCAN combines GCN, GAT, and GIN layers to capture hierarchical structural representations from 3D protein graphs, constructed from high-resolution PDB files with $\alpha$-carbon coordinates and residue types. To evaluate performance, we curated a high-quality dataset of 500 membrane and 500 non-membrane proteins, and compared MP-GCAN with two baselines: a structure-confidence-based SGD classifier utilizing AlphaFold’s pLDDT scores, and DeepTMHMM, a sequence-based deep learning model. Our experiments demonstrate that MP-GCAN significantly outperforms baselines, achieving an accuracy of 96\% and strong F1-scores on both classes. The results highlight the importance of integrating pretrained GNN architectures with domain-specific structural data to enhance membrane protein classification.

\end{abstract}


\IEEEpeerreviewmaketitle

\section{Introduction}

\IEEEPARstart{P}{rotein} classification is a crucial aspect of molecular biology, as proteins are the primary executors of cellular functions and their characteristics and interactions define the phenotypes and behaviors of cells. Recent advancements in proteomics and bioinformatics have significantly enhanced our understanding of the complex roles proteins play in various biological processes, including disease mechanisms such as cancer\cite{Khan2024cancer}. Membrane proteins are critical targets for drug development, with over 60\% of approved pharmaceuticals targeting their functions \cite{overington2006tapping}. Traditional methods for membrane protein prediction, such as transmembrane helix identification \cite{krogh2001predicting}, have laid the foundation for computational analysis of protein structures. The emergence of high-throughput technologies has revolutionized genome-wide investigations of protein expression and interaction networks, offering essential perspectives into the mechanisms underlying tumor development and cancer progression\cite{Men2021multi}. By integrating diverse layers of omics information, multi-omics-based approaches enhance the ability to detect nuanced yet biologically meaningful variations that may be overlooked by single-omics strategies. Such integrative analysis is essential for deepening our molecular understanding of disease processes and for informing the design of more precise and effective therapeutic strategies.
 
As fundamental macromolecules underlying various biological processes and cellular functions,  like   DNA transcription, transmembrane transport, catalytic reaction and tRNA charging,  protein's classification is becomming a crucial technique . 

In order to solve the task, people developed a series models and methods to increase the accuracy of machine recognition. RNN and Transformer are models based on aimno-acid-residue-sequecnce. They Models like 3D-CNNs (3-Dimensional Convolution Neural Network) \cite{JiS20123Dconv}, GNNs have emerged as powerful tools for protein structure analysis, enabling the integration of spatial and sequential features \cite{gligorijevic2021structure}.
GNN (Graph Neural Network) based on both structural information and sequential information. Comparing to simply sequence based information, 3D convolution and GNN are more comprehensive and effective when it comes to protein isoform identification. 3D-CNN embeds 3D coordinate into a 3D image and output a characteristic vector to classify.  A GNN underlay recognition mechanism is by embedding hand-engineered features into high dimensional  vectors. Another class of protein classification model is fusion models. 

Graph Neural Network is known to its ablility to extract relative spatial information and SO3-equivalance.  In the coordinate composed of any four $\alpha$-carbon atoms, each amino acid residue is equivalent of SO3 orthogonal groups. Their flaws are There emerges three types.

Despite the emergence of numerous models in the field of protein classification, there has been a notable absence of models utilizing pretrained Graph Neural Networks (GNNs) to predict specific proteins. Traditional approaches often fail to fully capture the unique characteristics of proteins due to their reliance on general datasets, which may not adequately represent the specific traits of interest. In contrast, GNNs, when pretrained on specialized datasets, can selectively enhance the extraction of distinctive features while suppressing those that are overly similar, thereby improving performance on specific protein classification tasks.

Recent advances in protein structure prediction, exemplified by AlphaFold \cite{jumper2021highly}, have revolutionized the field by providing accurate 3D models for previously uncharacterized proteins. There exists a few advances in membrane protein prediction have been introduced both structure-based and sequence-based approaches, such as a pLDDT-driven statistical model that leverages residue-level structural confidence scores, and DeepTMHMM, a state-of-the-art deep learning framework that performs end-to-end classification directly from protein sequences without requiring manual interpretation.

In our experiments, we focused on membrane proteins, a class of proteins with unique structural and functional properties. We trained a GNN based model called MP-GCAN using a dataset specifically curated for membrane proteins. This allowed the GNN to learn and prioritize the unique spatial and sequence features that are critical for membrane protein classification. We then compared the performance of our pretrained GNN model with that of a Recurrent Neural Network (RNN) model and a GNN model trained on a general dataset.

The results of our experiments demonstrated that the pretrained GNN model achieved significantly higher accuracy and overall performance compared to the other models. This suggests that the ability of the pretrained GNN to selectively focus on distinctive features of membrane proteins, while downplaying overly similar traits, provides a clear advantage in classification tasks. Our findings highlight the importance of using specialized datasets for pretraining models when targeting specific protein classes, as this approach can lead to more accurate and reliable classification outcomes.

\section{Related work}

Membrane proteins play essential roles in various cellular processes, acting as key mediators in signal transduction, substance transport, and energy conversion. They account for approximately 30\% of all protein-coding genes in the human genome, and over 60\% of approved drugs target membrane proteins, underscoring their central importance in pharmaceutical development. However, membrane protein research remains challenging: their hydrophobic nature hinders purification, and the dynamic conformations of transmembrane regions increase the difficulty of structural determination. As a result, membrane proteins constitute only about 1.5\% of entries in the Protein Data Bank (PDB).

Traditional approaches for membrane protein identification rely primarily on sequence-based heuristics. Early tools such as TMHMM  and Phobius employed hidden Markov models and signal peptide recognition to infer transmembrane helices from 1D amino acid sequences\cite{krogh2001predicting} \cite{kall2004combined}. While these methods achieved considerable success for $\alpha$ -helical membrane proteins (e.g., TMHMM reports an accuracy of up to 97\%), they performed poorly on $\beta$-barrel membrane proteins, with accuracies often below 60\%. Subsequently, machine learning models such as support vector machines (SVMs) and random forests incorporated sequence-derived features—such as amino acid composition and evolutionary conservation—achieving up to 82\% accuracy on datasets like PDBTM. In recent years, deep learning methods have emerged as promising alternatives. For example, DeepTMHMM , built on a residual convolutional neural network architecture, achieved 98.2\% accuracy for transmembrane helix prediction on the TMHMM 2.0 benchmark\cite{hallgren2022deeptmhmm}. Nonetheless, these approaches are limited by their reliance on sequence information alone, which hampers their ability to capture complex 3D structural and functional characteristics of membrane proteins.

The advent of AlphaFold has enabled widespread access to high-quality predicted protein structures, stimulating the development of structure-based membrane protein classification methods. Some studies transform protein structures into distance matrices or voxel grids and apply convolutional or recurrent neural networks for classification. However, such representations often lose spatial topology and incur high computational costs due to 3D convolutions. Other approaches utilize sequence embeddings from AlphaFold directly for downstream classification tasks. For instance, TAPE achieved an AUC of 0.91 on the CAFA3 dataset, demonstrating strong generalizability. More recently, graph neural networks (GNNs) have shown great potential for membrane protein analysis. GNN-based methods represent residues (or atoms) as nodes and spatial proximity as edges, thus preserving critical 3D topological features. MP-GCAN applied graph convolutional networks to protein structure analysis and demonstrated a 15\% accuracy improvement over sequence-based baselines in function prediction tasks by incorporating distance-aware message passing \cite{gligorijevic2021structure}. Equivariant GNNs (EGNN) introduced geometry-aware message functions, reducing geometric prediction errors by 30\% in molecular conformation generation compared to traditional GNNs\cite{satorras2021equivariant}. D-SCRIPT  integrated EdgeConv to fuse sequence and inter-residue distance features, achieving an interface residue prediction F1-score of 0.72 on a PDB-based dataset of membrane–nonmembrane protein interactions\cite{sledzieski2021d}.

In this work, we propose an end-to-end heterogeneous graph neural network architecture, ProteinGNN, designed to classify membrane proteins directly from structural data. Our model stacks multiple feature fusion layers: (i) a GCNConv layer to perform global graph convolution and capture long-range dependencies; (ii) a GATConv layer employing an attention mechanism to adaptively weight neighboring residues and emphasize functionally critical sites; and (iii) GINConv/EdgeConv layers to explicitly model inter-residue distance features via high-order graph isomorphism and edge convolutions. Furthermore, we apply dynamic geometric augmentation (random rotations and translations) during training to enhance robustness.

\section{Methodology}

\begin{figure*}[htbp]  
  \centering  
  \includegraphics[width=\textwidth]{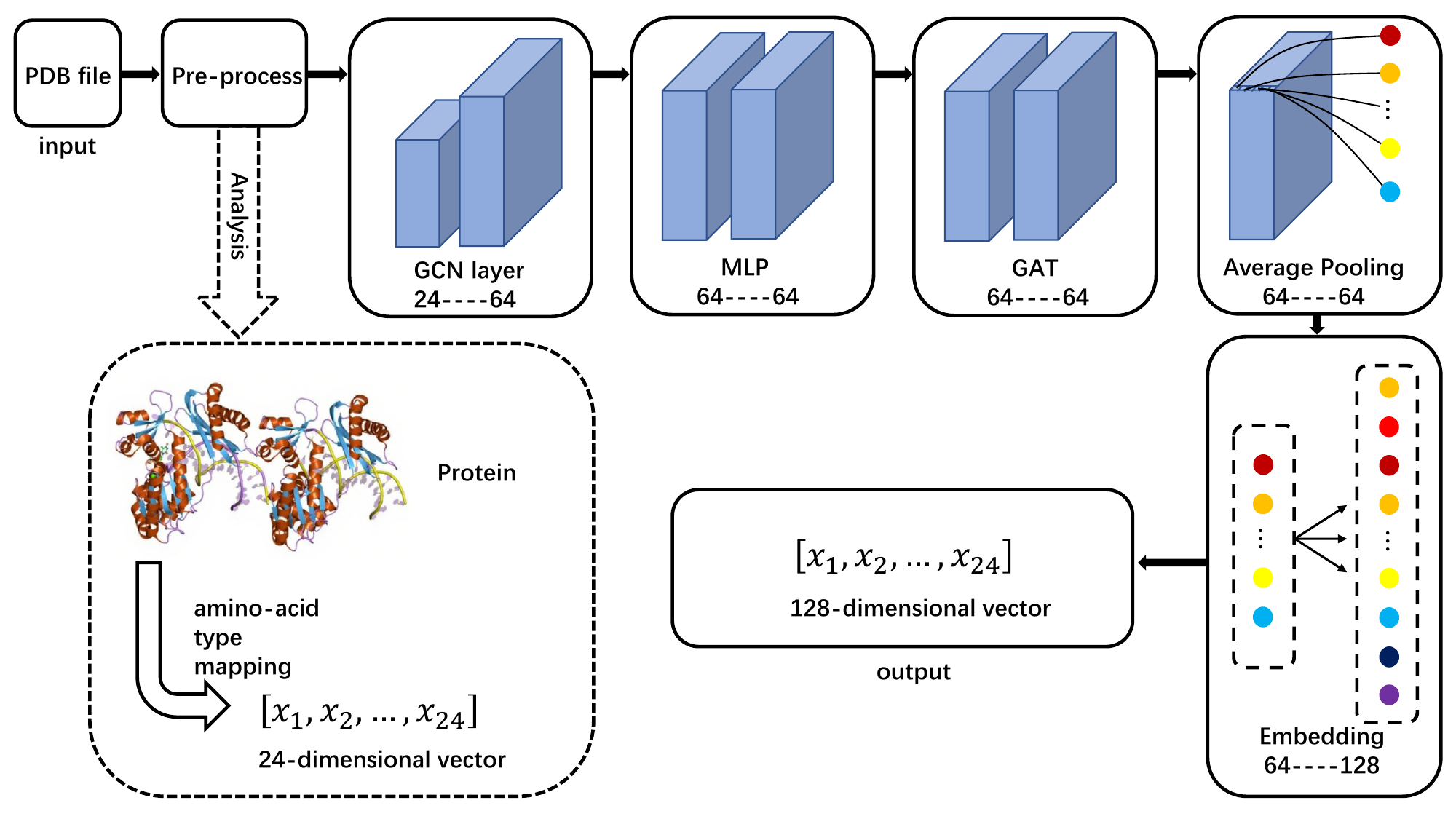}  
  \caption{FlowChart}  
  \label{fig:large_figure}  
  \raggedright \textit{The protein processing pipeline begins with a PDB file input that undergoes initial pre-processing before diverging into two parallel processing streams (Fig.~\ref{fig:large_figure}). The primary stream employs a series of graph-based neural network layers: a Graph Convolutional Network (GCN) layer first transforms the 24-dimensional input features into a 64-dimensional representation. This is followed sequentially by a Multilayer Perceptron (MLP) and Graph Attention Network (GAT) layer, both maintaining the 64-dimensional feature space. The architecture then applies average pooling before a final embedding layer projects the features into a 128-dimensional output vector.
  Concurrently, the secondary processing stream analyzes the pre-processed structural data through protein structure examination and amino-acid type mapping operations. This branch ultimately generates a complementary 24-dimensional feature vector $\mathbf{x} = (x_1, x_2, \dots, x_{24})$ that contributes to the final model output. The two streams collectively form an integrated framework that combines structural analysis with deep graph-based learning for comprehensive protein modeling.}

\end{figure*}

This section explains two different types of models we applied to classify membrane protein. There are two proposed methodology: Graph Neural Network based model and Recurrent Neural Network based model. The major difference between them is their data-processing-procedure. The baseline model simply extracts the sequential vector of each peptide chain. Graph Neural Network (GNN) can also process spatial information such as 3-dimensional coordinates and dihedral angle of $\alpha$-carbon atom.

\begin{subsection}{Graph Convolution Neural Network}
\subsubsection{\textbf{Overall Research Design}}
In this experiment, membrane protein datasets and non-membrane protein datasets are downloaded from RCSB Protein Bank in pdb format file format.
The model is consists of two graphic convolution layers and one self -attention-block with   parameters in total. After training with AdamW optimizer for 50 epochs, it came to the best model. Then similarity calculation and confusion matrix are applied to evaluate its performance.

The network is basically based on graph neural network structure. Original GNNs can capture not just spatial coordinates but also the local connection between nodes. Unlike CNNs limited to Euclidean data, GNNs efficiently extract non-Euclidean spatial correlations by hierarchically fusing neighbor embeddings within localized graph regions. That is why it is crucial in processing biomacromolecules' structure. MP-GCAN is composed of one data pre-processing layer, two graph convolution layers, one multi-perceptron-layer, one graph self attention layer, one average pool and one embedding layer.


\subsubsection{\textbf{Data Pre-processing Layer}}
Data sets are preserved in three file folders: train set, validation set and test set. First, we read peptides chain's residue sequence and map them into a 21-dimensional-vector through a global residue function.
\begin{align}
    \mathrm{GR}: \{\text{amino acid type}\} & \rightarrow \{z \in \mathbb{Z}\} 
\end{align}
There emerges 21 types of amino acid in our datasets. (Including 20 types of standard amino acid and 'UNK' amino acid). Then, $\alpha$-carbon coordinates are concatenated to the vector's end and generate the feature vector of node. Edge relations between nodes are represented by self-loop-included-matrix with $a_{ij}$ = \{1, if node $i$ and $j$ are directly connected and their Euclidean distance is less than 6\,\AA; 0, otherwise\}.\\
\[
A = \begin{bmatrix}
a_{11} & \cdots & a_{1n} \\
\vdots & \ddots & \vdots \\
a_{m1} & \cdots & a_{mn}
\end{bmatrix}
\]

\subsubsection{\textbf{Graph Convolution Layer}}
There are two multi-channel-convolution-layers in total. Let $G=(V,E)$ be the graph structure. Adjacency matrix is represented by $A \in GL(\mathbb{R}^N)$, where $N=|V|$ represents the number of nodes. Assume $X \in GL(\mathbb{R}^{N\times C_{in}})$ to be the node feature matrix, where $C_{in}$ denotes the count of input channels, i.e. dimension of input vectors. $H^{i} \in \mathbb{R}^{C_{in}}$ is the $i$-th channel of output layer. Each layer takes the adjacency matrix and node feature matrix as input and outputs a new feature matrix $H \in GL(\mathbb{R}^{N \times C_{out}})$ satisfying \cite{zit2018modgcn}:
\begin{align}
    H^i = \sigma (D^{-\frac{1}{2}}AD^{-\frac{1}{2}} X W^i).
\end{align}
$\sigma = ReLU$ is convolution layers' activation function. $D \in GL(\mathbb{R}^{N})$ ($D_{ii} = \sum_j a_{ij}$, $D_{ij} = 0$ while $i \neq j$) is used to normalize the adjacency matrix. $W \in GL(\mathbb{R}^{C_{in} \times C_{out}})$ and $W^i$ is the $i$-th column vector of $W$. They are learnable parameters. Each output channel corresponds to a weight vector $W^i$.

Graph convolution layers are applied to retrieve topological message. Via message passing, they aggregate neighborhood information to capture both fine-grained local geometry and global spatial patterns. Compared to 3D CNNs, GNNs are more effective in handling sparse and irregular structures. These properties make GNNs particularly powerful for tasks such as protein classification.

\subsubsection{\textbf{Multiple Perceptron Layer}}
In MP-GCAN, there exists two fully connected-layers in MLP. Each layer has 64 neurons. $X$ is node feature matrix put in MLP. $Z \in \mathbb{R}^{64}$ is a shared weight vector of 64 dimensions that applied to feature vector $X^i$ of every node. And it is a trainable parameters' set. During the training procedure, it will be trained with drop-out rate of $0.2$ in order to enhancing generalization ability and robustness.

Node feature matrix is the only input of MLP. Edge index and adjacency matrix remain the same during the process. It is utilized to reconstruct vectors' spatial features and project them into a easier classified space. By using consistent dimension of 64, MLP efficiently avoid information compression. It is also applied as a transition layer between GCNs and GAT \cite{10606483}. MLP also serves to increase non-linearity and lead to enhanced class distinguishability.

\subsubsection{\textbf{Graph self Attention Layer}}
There exist a single head attention layer after MLP. Assume $X \in GL(\mathbb{R}^{N \times F_{in}})$ to be node feature matrix, $\mathbf{X} = \begin{bmatrix} X_1 & X_2 & \cdots & X_n \end{bmatrix}^{T}$, where $F_{in}$ is the input feature dimension. Afterwards, they are projected to a $F_{out}$ dimensional space with $W \in GL(\mathbb{R}^{F_{out} \times F_{in}})$.
\begin{align}
    \mathbf{z_i} = WX_i.
\end{align}
$W$ is a trainable weight matrix. Next, for each pair of connected nodes $(i,j)$, an attention score $e_{ij}$ is computed using a learnable weight vector $\mathbf{a} \in \mathbb{R}^{2F_{out}}$, as follows:
\begin{align}
    e_{ij} = LeakyReLU(\mathbf{a}^T [\mathbf{z_i}||\mathbf{z_j}]).
\end{align}
$[\mathbf{z_i}||\mathbf{z_j}]$ denotes the concatenation of the two transformed feature vectors with $LeakyReLU$ as activation function ($\alpha$ = 0.2). The score $e_{ij}$ on behalf of the importance of neighbor $j$ to node $i$. To normalize the attention scores, softmax function is applied over all $j$ in $i$'s neighborhood.
\begin{align}
    \alpha_{ij} = \frac{exp(e_{ij})}{\sum_{k \in \mathcal{N}(i)}exp(e_{ik})}.
\end{align}
The resulting attention coefficient $\alpha_{ij}$ quantifies the relative importance of neighbor $j$ among all neighbors of node $i$. In the end, the output feature of $i$-th node ($\mathbf{h}_i$) is:
\begin{align}
    \mathbf{h_i} = \sigma(\sum_{j \in \mathcal{N}(i)} \alpha_{ij}X_j).
\end{align}

\subsubsection{\textbf{Average Pool Layer \& Embedding Layer}}
In the end, average pool layer and embedding layer are employed to adjust output vectors' length. Without loosing vectors spatial feature, they play a role of distinguishing few clusters of protein better.
\end{subsection}

\subsection{baseline model}
\subsubsection{\textbf{pLDDT}}
To directly validate structural uncertainty, we designed a new baseline model that relies only on structural confidence rather than sequence information. Membrane proteins exhibit different structural confidence characteristics from non-membrane proteins due to their stable transmembrane domains. The model takes the predicted 3D structure in PDB format as input, extracts the pLDDT score for each residue, and calculates a statistical feature vector including the mean, median, standard deviation, and the percentage of residues with a score below 70. These features are processed using pre-trained interpolators  and scalers to handle missing values and normalize the data, respectively. Finally, a stochastic gradient descent (SGD) classifier  is trained on these feature vectors to predict whether the protein is "membrane" or "non-membrane".

\subsubsection{\textbf{DeepTMHMM}}
DeepTMHMM is a modern sequence-based, state-of-the-art (SOTA) membrane protein prediction method that leverages advanced deep learning techniques. Unlike traditional methods, it uses a deep neural network architecture to automatically extract information directly from amino acid sequences. The input consists of protein sequences in FASTA format, similar to the classic TMHMM, but DeepTMHMM is unique in that it provides an end-to-end prediction pipeline. It does not require human interpretation, but directly outputs classification labels, which simplifies the prediction process by eliminating the need for additional post-processing.

\section{Experiments}

\subsection{Datasets}
    
For this study, we build a dataset with two high-quality groups to compare membrane proteins and non-membrane proteins, each containing 500 structures. The membrane protein structures come from the OPM database (\url{https://opm.phar.umich.edu/}), which is known for reliable annotations of transmembrane regions and how proteins sit in membranes. For non-membrane proteins, we use structures from the PDB (\url{https://www.rcsb.org/}), the main repository for experimentally determined protein structures, ensuring the data collection timeline matches the membrane protein set.

To maintain high quality, both groups undergo strict checks. We include only X-ray structures with resolution $\leq 2.5$\AA\ and cryo-EM structures with global resolution $<3.0$\AA, excluding NMR structures due to their lower atomic precision. We also remove structures with more than 5\% missing backbone residues (excluding known disordered regions) and exclude engineered mutants, fusion proteins, or those with non-physiological ligands.

For membrane proteins, we prioritize $\alpha$-helical polytopic types---specifically including only those with at least 2 transmembrane $\alpha$-helices. We also retain representatives from other membrane-associated subclasses, such as peripheral membrane proteins, lipid-anchored proteins, and $\beta$-barrel types, ensuring coverage of major membrane protein architectures.

For non-membrane proteins, we ensure the exclusion of any membrane-associated ones: structures labeled as membrane proteins in PDB or predicted to have transmembrane helices are removed. This group includes proteins from different functional classes like enzymes, antibodies, and transcription factors to reflect the diversity of non-membrane proteins.

\subsection{Model Hyperparameters}

\begin{table}[t]
\caption{Model Hyperparameters}
\label{tab:hyperparams}
\centering
\footnotesize
\begin{tabular}{@{}ll@{}}

\textbf{Parameter} & \textbf{Value} \\

Input dimension & 24 (21 AA types + 3D coords) \\
Hidden dimension & 64 \\
Embedding dimension & 128 \\
GNN layers & GCN $\rightarrow$ GAT $\rightarrow$ GIN \\
Edge threshold & {6.0}{\angstrom} \\
Batch size & 8 \\
Epochs & 10 \\
Learning rate & 0.001 (Adam) \\
Dropout rates & 0.2 (GNN), 0.3 (classifier) \\

\end{tabular}
\end{table}

We develop the MP-GCAN architecture using carefully selected hyperparameters (Table~\ref{tab:hyperparams}) that balance model complexity with computational efficiency. The network takes as input 24-dimensional node features that encode both amino acid identity (21 dimensions for standard residues) and 3D spatial coordinates. These features pass through three specialized graph neural network layers --- first a graph convolution (GCN), then a graph attention layer (GAT), and finally a graph isomorphism layer (GIN) --- each with 64 hidden units to capture hierarchical protein structural patterns.

Following common practice in protein structure analysis, we connect amino acids (via their C$\alpha$ atoms) if they were within 6.0{\angstrom} of each other, creating edges that represent potential biological interactions. During training, we used the AdamW optimizer with a learning rate of 0.001 and applied dropout (rates of 0.2--0.3) to different network components to prevent overfitting to the training data. The model typically converged within 10 training epochs using modest batch sizes of 8 protein structures at a time. This configuration produced 128-dimensional protein embeddings that effectively distinguished between our target classes in subsequent analysis.

\subsection{Baseline}

\textbf{pLDDT-based Structural Confidence Classifier.} 
We designed this novel baseline to test our hypothesis that structural uncertainty patterns discriminate membrane proteins. Using only per-residue pLDDT confidence scores from AlphaFold-predicted structures (no sequence information), we extracted statistical features (mean, median, standard deviation, and percentage of low-confidence residues) which were normalized and fed into a Stochastic Gradient Descent (SGD) classifier. This approach intentionally isolates structural confidence as the sole predictive factor.

\textbf{DeepTMHMM} 
While TMHMM 2.0 and DeepTMHMM represent important benchmarks in membrane protein prediction, we note these models were trained on different datasets than our ProteinGNN framework. The conventional TMHMM 2.0 utilizes proprietary training data unavailable for our study, while DeepTMHMM's implementation details and training corpus remain undisclosed. Due to this fundamental difference in training conditions - where our baselines weren't exposed to the same distribution of membrane/non-membrane proteins during their development - we refrain from presenting direct performance comparisons. Such unequal comparisons could yield misleading conclusions about relative model capabilities. Instead, we focus our quantitative analysis on the pLDDT-based classifier which, like our MP-GCAN, was trained exclusively on our curated dataset under identical conditions.

Crucially, all models were evaluated on the same independently curated test set using identical metrics (accuracy, F1-score, etc.). This controlled comparison isolates methodological differences as the sole variable when interpreting performance gaps between our MP-GCAN (which fuses structural and sequence data) and these baselines representing structural-only, classical sequence, and modern deep learning paradigms.

\subsection{Training}
\subsubsection{\textbf{Loss Function}}
Cross-entropy is a widely used criterion for multi-class classification tasks, as it effectively measures the divergence between the predicted probability distribution and the true label distribution. In our case, the model outputs a softmax-normalized probability vector for each protein graph, and the loss penalizes incorrect predictions in proportion to their confidence.  Additionally,  an L2 regularization term was appended to increase MP-GCAN's robustness.

\subsubsection{\textbf{Optimizer}}
During the training process, we employ the AdamW optimizer\cite{AdamW} for its robustness in adapting per-parameter learning rates, which is particularly beneficial for sparse graph structures.

\subsubsection{\textbf{Regularization}}
During optimization, the L2 penalty is applied to gradient updates, promoting smaller weights and robust feature representation. Combined with dropout regularization at multiple levels (dropout rate = $0.2$), MP-GCAN achieves stable convergence and improved generalization.

\section{Results}

\begin{figure}[htbp]
    \centering
    \setlength{\abovecaptionskip}{4pt}  
    \setlength{\belowcaptionskip}{2pt}
    
    \subfloat[Loss and accuracy curve\label{subfig:curves2}]{%
        \includegraphics[width=0.95\columnwidth,height=3cm]{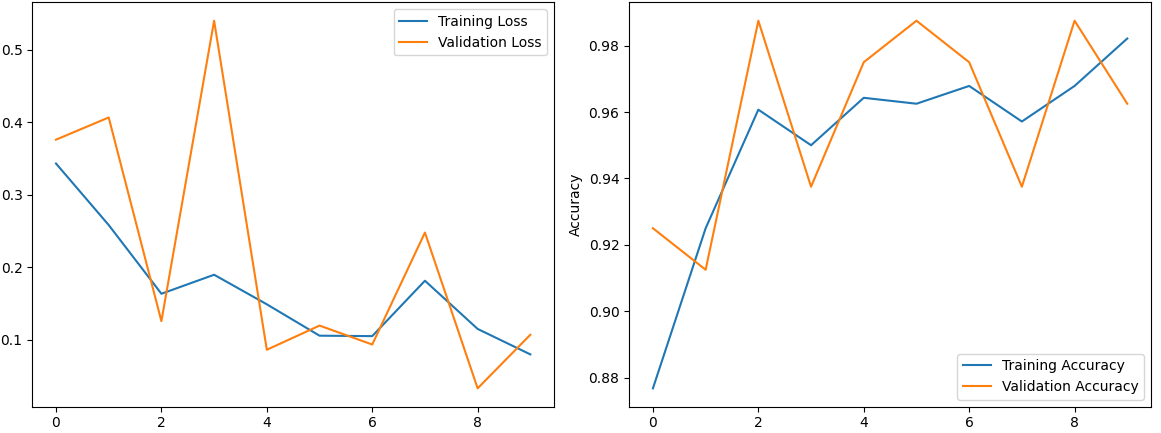}
    }
    
    \vspace{0.5cm}
    
    \subfloat[t-SNE Protein embedding distribution\label{subfig:tsne2}]{%
        \includegraphics[width=0.95\columnwidth,height=6cm]{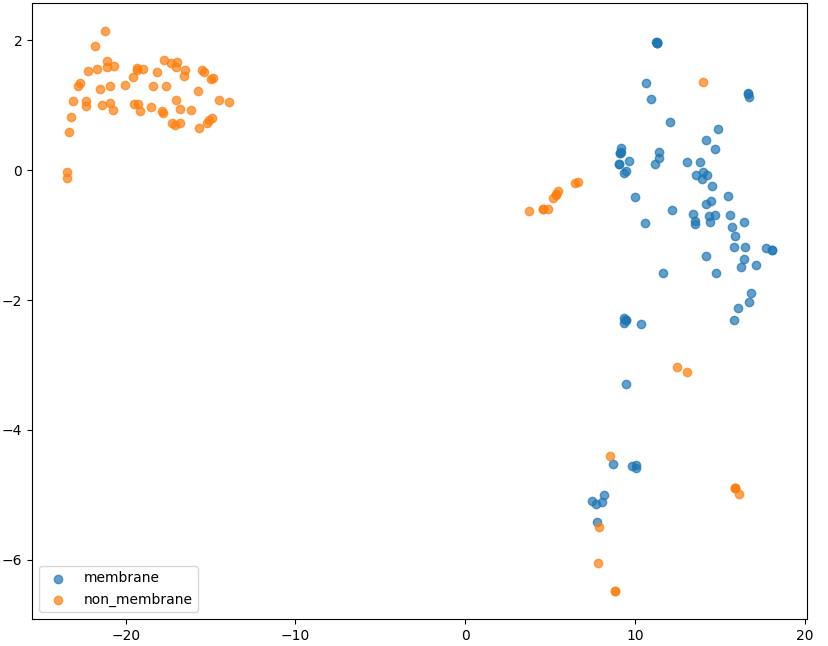}
    }
    
    \caption{$\alpha$-helical proteins and non-membrane proteins key training visualizations}
    \label{fig:training_vis2}
\end{figure}

\subsection{Result:$\alpha$-helical proteins and non-membrane proteins}
The t-SNE visualization reveals distinct clustering patterns between $\alpha$-helical membrane proteins and soluble proteins in the embedding space, demonstrating our model's ability to capture key structural discriminants. As shown in Fig.~\ref{subfig:tsne2},the majority of soluble proteins form a well-separated cluster, the presence of overlapping regions suggests certain structural ambiguities in the dataset. These transitional zones likely correspond to membrane-interacting proteins or soluble domains with structural similarities to membrane helices, reflecting the continuous nature of structural space rather than discrete categories. The clear separation achieved for most cases validates our approach, while the borderline instances highlight biologically meaningful edge cases that warrant further investigation.

During the training process, the loss value steadily decreased (to 0.1 in the 9th round), and the training accuracy was as high as 98\%, indicating that the model has fully learned the hydrophobic pattern and length distribution of transmembrane helices. However, the validation loss fluctuated in the 3rd/7th round, and the accuracy decreased synchronously, which was caused by the complexity of the samples. This shows that although the model has strong generalization ability, its ability to learn edge cases of complex proteins is still insufficient.

\begin{table}[!t]
\centering
\caption{Comparison of Model Performance}
\label{tab:model_performance}
\begin{threeparttable}

    \begin{tabular}{|l|l|c|c|c|}
    \hline
    \textbf{Model} & \textbf{Category}    & \textbf{Recall} & \textbf{F1-Score} & \textbf{Accuracy} \\ 
    \hline
    \multirow{2}{*}{pLDDT} 
                  & Membrane            & 0.82            & 0.77              & \multirow{2}{*}{0.75} \\ \cline{2-4}
                  & Non-membrane        & 0.68            & 0.73              & \\ \hline
   
    \multirow{2}{*}{MP-GCAN} 
                  & Membrane            & 0.93            & 0.96              & \multirow{2}{*}{0.96} \\ \cline{2-4}
                  & Non-membrane        & 0.92            & 0.96              & \\ \hline
    \end{tabular}

\begin{tablenotes}
\small
\item[1] \textit{This table compares the performance of pLDDT and MP-GCAN models on membrane and non-membrane protein classification tasks. } 
\item[2] \textit{Recall measures the ability to correctly identify all relevant samples. F1-Score is the harmonic mean of precision and recall, balancing both metrics. Accuracy reflects the overall correctness of predictions.  }
\end{tablenotes}
\end{threeparttable}
\end{table}

From Table n, we can see the performance comparison. Because it is difficult to model the spatial folding of transmembrane helices and the dynamic interaction of the membrane environment, the traditional method pLDDT has a low recall rate (0.82) and F1 score (0.77) on $\alpha$-helical multi-spanning membrane proteins. However, MP-GCAN improves the recall rate to 0.93 and F1 to 0.96 through the graph structure. Its breakthrough lies in accurately capturing subtle features such as helical bending and transmembrane interactions, which significantly optimizes the recognition ability of multi-spanning membrane topology.

\subsection{Result:Add Beta-barrel transmembrane}

\begin{figure}[htbp]
    \centering
    \setlength{\abovecaptionskip}{4pt}  
    \setlength{\belowcaptionskip}{2pt}
    
    \subfloat[Loss and accuracy curve\label{subfig:curves3}]{%
        \includegraphics[width=0.95\columnwidth,height=3cm]{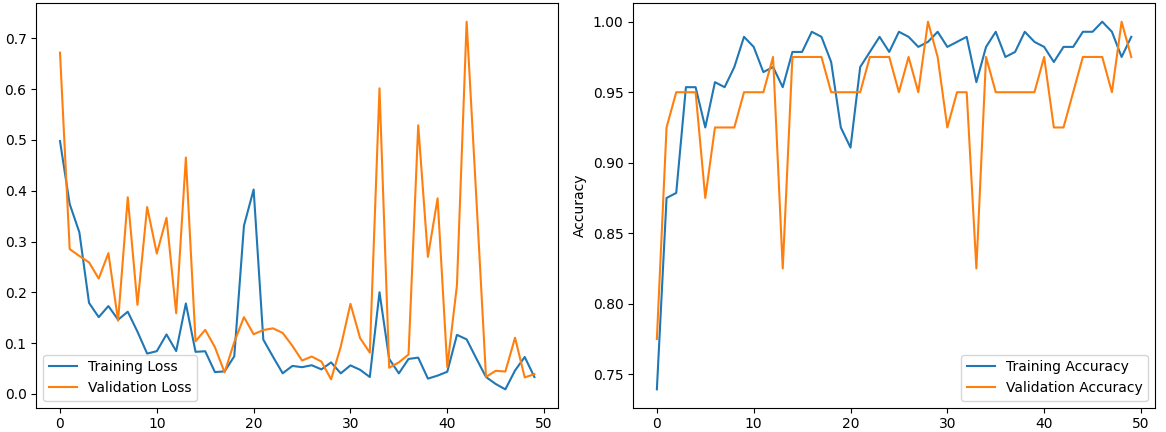}
    }
    
    \vspace{0.5cm}
    
    \subfloat[t-SNE Protein embedding distribution\label{subfig:tsne3}]{%
        \includegraphics[width=0.95\columnwidth,height=6cm]{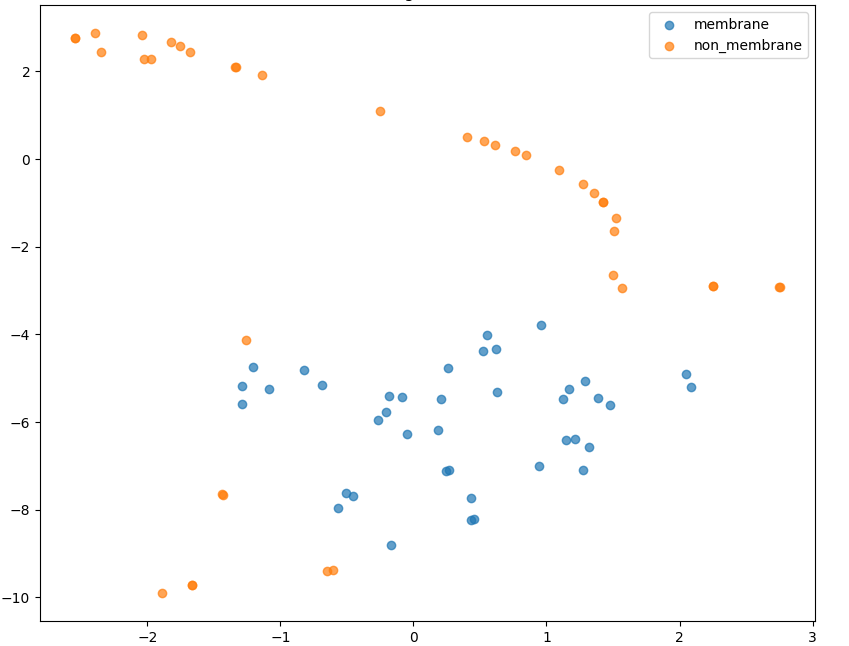}
    }
    
    \caption{Adding key training visualizations of $\beta$ proteins}
    \label{fig:training_vis3}
\end{figure}

In order to further verify the generalization ability of the model, we changed the original dataset to Beta-barrel transmembrane and increased the number of training rounds to 50.

From Fig.~\ref{subfig:tsne3}  we can see that the training and validation curves over 50 epochs demonstrate the model's robust learning capability when handling the structurally diverse dataset containing both $\alpha$-helical and $\beta$-barrel membrane proteins. The training loss consistently decreases and eventually stabilizes around 0.05, indicating effective feature learning across different membrane protein topologies. While the validation loss shows occasional spikes at specific epochs (e.g. 10 and 35), corresponding to challenging edge cases like low-sequence-homology $\beta$-barrels or hybrid-topology proteins, these fluctuations are successfully mitigated as the model's generalization capability improves. The training accuracy rapidly exceeds 95\% after just 10 epochs and maintains this high level, while the minimal gap ($<$5\%) between training and validation accuracy confirms the model's strong generalization performance for real-world applications involving complex membrane protein mixtures.

The t-SNE visualization of 128-dimensional embeddings reveals biologically meaningful clustering patterns that validate the model's discriminative power. Membrane proteins naturally separate into two distinct subclusters: a dense central grouping representing $\alpha$-helical bundles with their characteristic packing geometries, and a more dispersed lower-left region corresponding to $\beta$-barrel proteins with their unique curvature signatures. Notably, soluble proteins maintain clear separation despite the increased complexity from $\beta$-barrel inclusion, demonstrating the model's focused learning of membrane-specific structural features such as lipid-facing residue patterns and absolute solvent accessibility.

\subsection{Result:Three classification problems}
    
\begin{figure}[htbp]
    \centering
    \setlength{\abovecaptionskip}{4pt}  
    \setlength{\belowcaptionskip}{2pt}
    
    \subfloat[pLDDT\label{subfig:curves4}]{%
        \includegraphics[width=0.95\columnwidth,height=7cm]{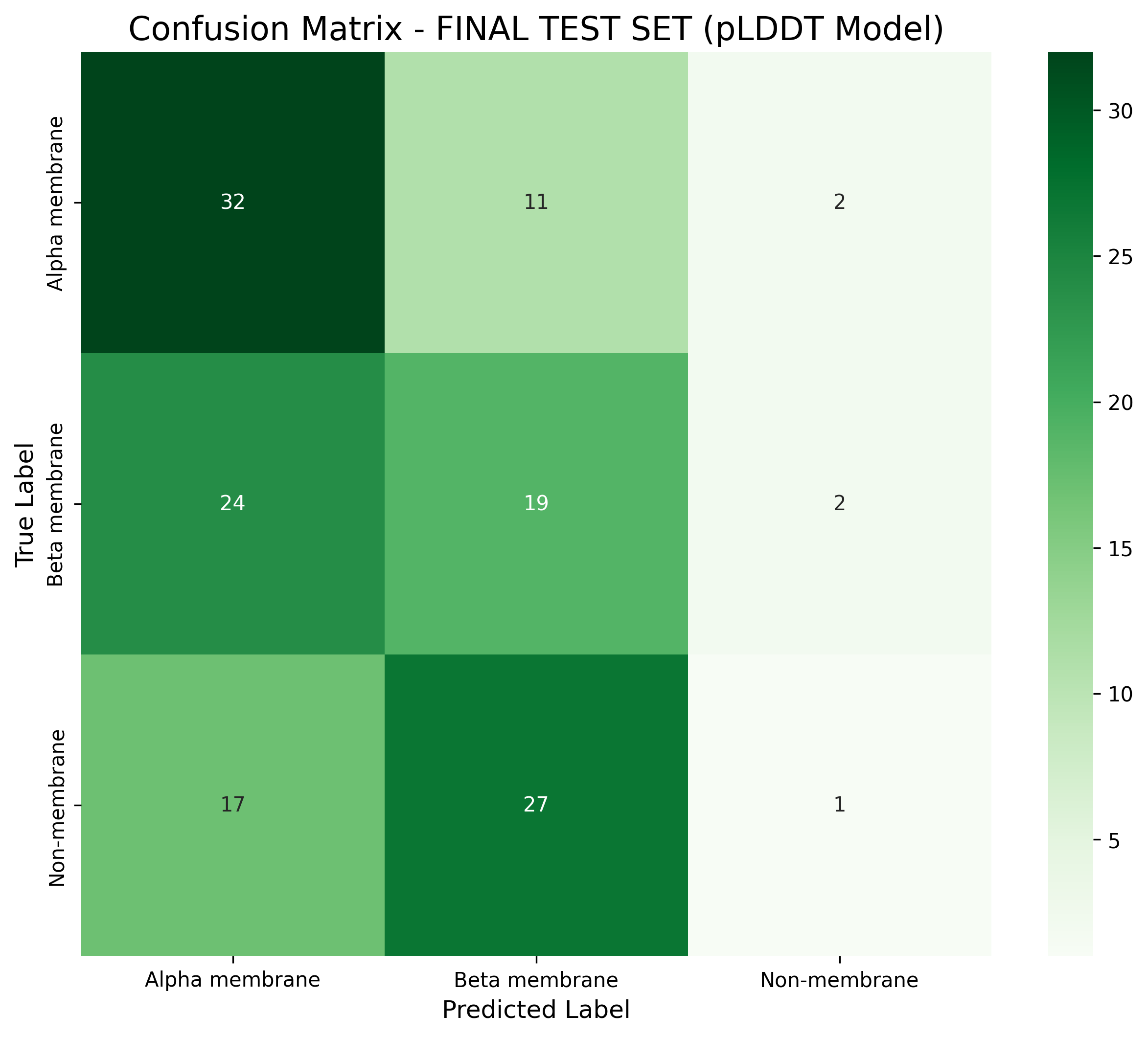}
    }
    
    \vspace{0.5cm}
    
    \subfloat[DeepTMHMM\label{subfig:tsne4}]{%
        \includegraphics[width=0.95\columnwidth,height=7cm]{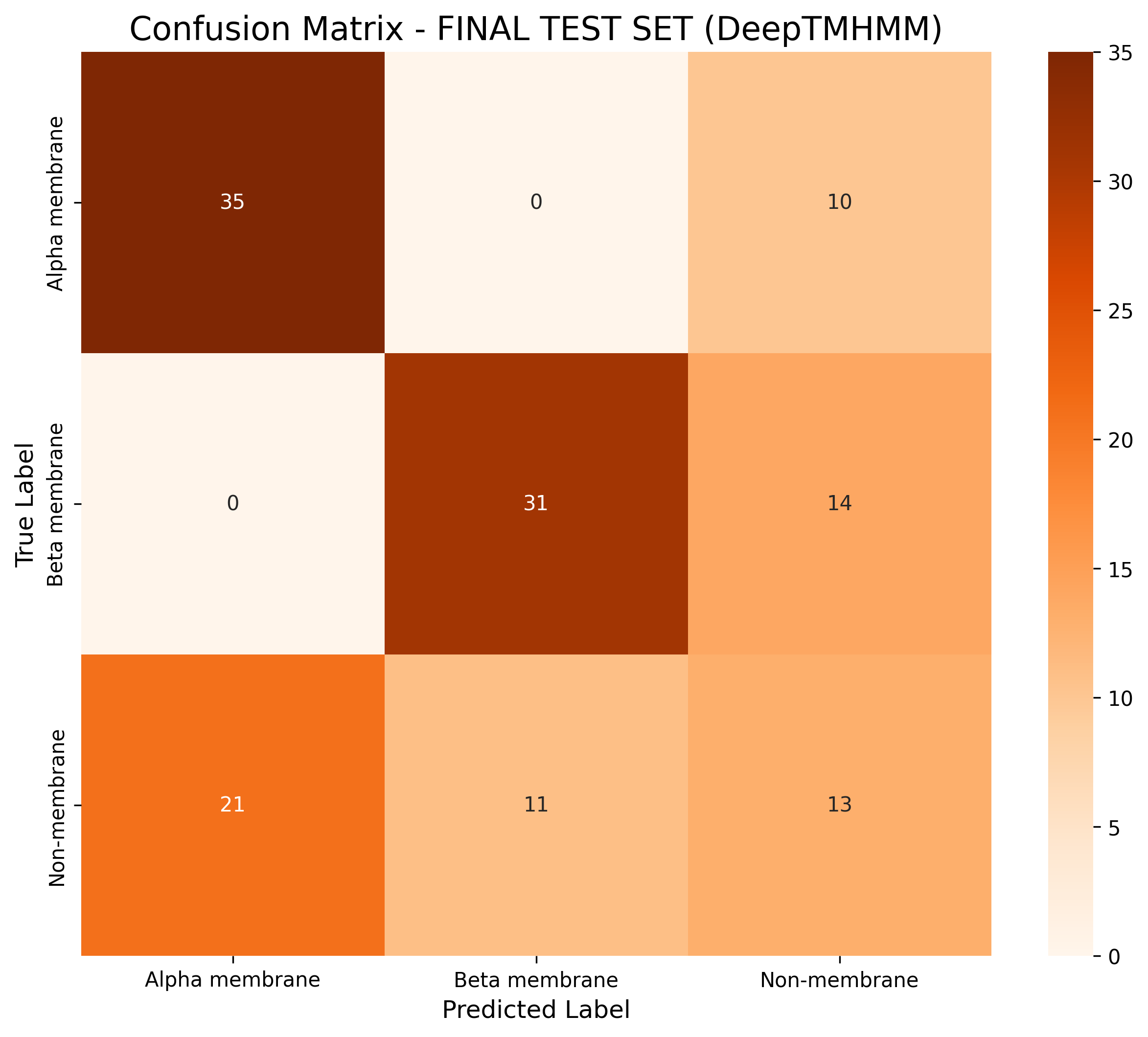}
    }
    
    \vspace{0.5cm}
    
    \subfloat[MP-GCAN\label{subfig4}]{%
        \includegraphics[width=0.95\columnwidth,height=7cm]{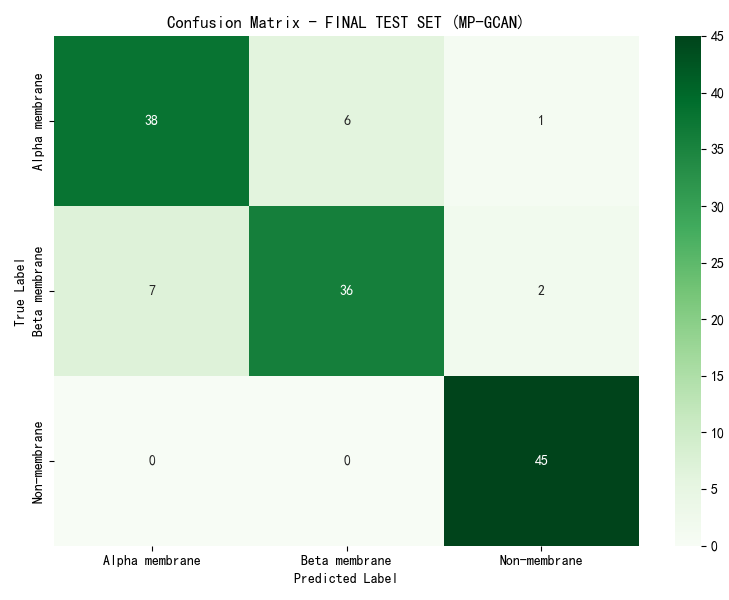}
    }
    
    \caption{Baseline three-class confusion matrix}
    \label{fig:training_vis4}
\end{figure}

\begin{table}[htbp]
    \centering
    \caption{Model Performance Summary}
    \label{tab:model_metrics}
    \begin{tabular}{|l|l|c|c|c|}
    \hline
    \textbf{Model} & \textbf{Category} & \textbf{Recall} & \textbf{F1-Score} & \textbf{Accuracy} \\ \hline
  
    \multirow{3}{*}{MP-GCAN} 
                  & $\alpha$-helical & 0.84 & 0.84 & \multirow{3}{*}{0.893} \\ 
    \cline{2-4}
                  & $\beta$-barrel & 0.84 & 0.85 & \\ 
    \cline{2-4}
                  & Non-membrane & 1.00 & 0.97 & \\ 
    \hline

    \multirow{3}{*}{pLDDT Model} 
                  & $\alpha$-helical & 0.71 & 0.60 & \multirow{3}{*}{0.622} \\ 
    \cline{2-4}
                  & $\beta$-barrel & 0.40 & 0.36 & \\ 
    \cline{2-4}
                  & Non-membrane & 0.20 & 0.20 & \\ 
    \hline

    \multirow{3}{*}{DeepTMHMM} 
                  & $\alpha$-helical & 0.78 & 0.69 & \multirow{3}{*}{0.690} \\ 
    \cline{2-4}
                  & $\beta$-barrel & 0.69 & 0.70 & \\ 
    \cline{2-4}
                  & Non-membrane & 0.34 & 0.34 & \\ 
    \hline
    \end{tabular}
\end{table}

To systematically evaluate model performance, we conducted a rigorous three-way classification experiment distinguishing $\alpha$-helical membrane proteins, $\beta$-barrel membrane proteins, and non-membrane proteins. Our comparative analysis examined three distinct computational approaches: the structure-aware graph neural network MP-GCAN, the confidence-based pLDDT Model, and the sequence-based DeepTMHMM method.

In the three-class classification task distinguishing $\alpha$-helical membrane proteins, $\beta$-barrel membrane proteins, and non-membrane proteins, MP-GCAN outperformed the pLDDT Model and DeepTMHMM, highlighting the advantages of structure-aware graph neural networks (Table~\ref{tab:model_metrics}).

MP-GCAN achieved the highest overall accuracy (0.893) with strong performance across all classes: For $\alpha$-helical membrane proteins, it showed balanced precision, recall, and F1-score (0.844), effectively capturing helical transmembrane features. For $\beta$-barrel membrane proteins, it reached a precision of 0.857, recall of 0.837, and F1-score of 0.847, successfully identifying the unique curvature of $\beta$-barrel structures. Notably, for non-membrane proteins, MP-GCAN achieved near-perfect recall (1.000) and high precision (0.938), with an F1-score of 0.968, demonstrating superior ability to distinguish soluble proteins.

In contrast, the pLDDT Model performed poorly (accuracy = 0.622), especially for non-membrane proteins (F1 = 0.200), due to its reliance on structural confidence alone. DeepTMHMM showed moderate performance (accuracy = 0.690) but failed to classify non-membrane proteins effectively (F1 = 0.342), reflecting limitations of sequence-based approaches in capturing 3D structural differences.

\section{Conclusion}
This study presents an application of GNN based model over membrane protein classification. MP-GCAN mainly concatenates two GCN layers, an MLP with two fully connected layers, and a GAT layer to extract protein's sequential and spatial feature to be a series of vectors and send them to a classifier. 

This study demonstrates that MP-GCAN, a graph neural network integrating sequence and 3D structural features, outperforms baseline models in three-class membrane protein classification. By leveraging graph convolutions and attention mechanisms, it effectively captures spatial topology and residue interactions, addressing key limitations of sequence-based or structural-confidence-only methods.
Membrane proteins are critical for cellular processes and drug targeting (over 60\% of approved drugs target them), making accurate classification vital for biological research and therapeutic development. MP-GCAN’s high performance highlights the value of structure-aware deep learning in protein analysis, with potential to advance functional annotation and drug discovery for membrane proteins.

\bibliographystyle{IEEEtran}  
\bibliography{references}     

\end{document}